\documentstyle[amssymb,psfig]{mn}

\newcommand{\Msun}{\mbox{$M_{\odot}$}}
\newcommand{\Mi}{\mbox{$M_i$}}
\newcommand{\tdis}{\mbox{$t_{\rm dis}$}}
\newcommand{\tdisref}{\mbox{$t_4^{\rm dis}$}}
\newcommand{\tburst}{\mbox{$t_{\rm burst}$}}

\title[Star Cluster Formation and Disruption Time-Scales -- II.]{Star
Cluster Formation and Disruption Time-Scales -- II.  Evolution of the
Star Cluster System in M82's Fossil Starburst}

\author[R.  de Grijs, N. Bastian, and H.J.G.L.M.  Lamers]{Richard de
Grijs$^1$\thanks{E-mail: grijs@ast.cam.ac.uk}, Nate Bastian$^2$, and 
Henny J.G.L.M. Lamers$^2$
\\ 
$^1$ Institute of Astronomy, University of Cambridge, Madingley Road,
Cambridge CB3 0HA\\
$^2$ Astronomical Institute, Utrecht University, Princetonplein 5,
3584 CC Utrecht, The Netherlands
}

\date{Received date; accepted date}
\pubyear{2001}

\begin{document}
\maketitle

\begin{abstract}
We obtain new age and mass estimates for the star clusters in M82's
fossil starburst region B, based on improved fitting methods.  Our new
age estimates confirm the peak in the age histogram attributed to the
last tidal encounter with M81; we find a peak formation epoch at
slightly older ages than previously published, $\log( t_{\rm peak} /
{\rm yr} ) = 9.04$, with a Gaussian $\sigma$ of $\Delta \log( t_{\rm
width}) = 0.273$.  The actual duration of the burst of cluster formation
may have been shorter because uncertainties in the age determinations
may have broadened the peak.  Our improved mass estimates confirm that
the (initial) masses of the M82 B clusters with $V \le 22.5$ mag are
mostly in the range $10^4 - 10^6 M_\odot$, with a median mass of $M_{\rm
cl} = 1.08 \times 10^5 M_\odot$.  The formation history of the observed
clusters shows a steady decrease towards older ages.  This indicates
that cluster disruption has removed a large fraction of the older
clusters. \\
Adopting the expression for the cluster disruption time-scale of
$\tdis(M)= \tdisref (M/10^4 \Msun)^{\gamma}$ with $\gamma \simeq 0.62$
(Paper I), we find that the ratios between the real cluster formation
rates in the pre-burst phase ($\log(t/{\rm yr})\ge 9.4$), the
burst-phase ($8.4 < \log(t/{\rm yr}) < 9.4$) and the post-burst phase
($\log(t/{\rm yr}) \le 8.4$) are about $1:2:{1\over 40}$.  The formation
rate during the burst may have been higher if the actual duration of the
burst was shorter than adopted. \\
The mass distribution of the clusters formed during the burst shows a
turnover at $\log(M_{\rm cl}/M_\odot) \simeq 5.3$ which is not caused by
selection effects.  This distribution can be explained by cluster
formation with an initial power-law mass function of slope $\alpha=2$ up
to a maximum cluster mass of $M_{\rm max}=3 \times 10^6 \Msun$, and
cluster disruption with a normalisation time-scale $\tdisref /
\tburst=(3.0 \pm 0.3) \times 10^{-2}$.  For a burst age of $1 \times
10^9$ yr, we find that the disruption time-scale of a cluster of $10^4$
\Msun\ is $\tdisref \sim 3 \times 10^7$ years, with an uncertainty of
approximately a factor of two.  This is the shortest disruption
time-scale known in any galaxy. 
\end{abstract}

\begin{keywords}
diffusion -- galaxies: individual: M82 -- galaxies: starburst --
galaxies: star clusters
\end{keywords}

\section{Introduction}
\label{intro.sec}

\subsection{Multiple starbursts in M82}

M82 is often regarded as the ``prototype'' starburst galaxy. 
Observations over the entire wavelength range, from radio waves to
X-rays (see, e.g., Telesco 1988 and Rieke et al.  1993 for reviews),
seem to support a scenario in which tidal interactions, predominantly
with its large neighbour M81, triggered intense star formation in the
centre of this small irregular galaxy during the last several 100 Myr. 
The resulting starburst, with the high star formation rate of $\sim 10
M_\odot$ yr$^{-1}$, has continued up to about 50 Myr ago (e.g.,
O'Connell \& Mangano 1978, Rieke et al.  1993).  Energy and gas ejection
from supernovae and combined stellar winds drive a large-scale galactic
wind along the minor axis of M82 (e.g., Lynds \& Sandage 1963, McCarthy
et al.  1987, Shopbell \& Bland-Hawthorn 1998). 

All of the bright radio and infrared sources associated with the active
starburst are found in a small region within a radius of $\sim 250$ pc
from the galaxy's centre, but most of this volume is heavily obscured by
dust at optical wavelengths. 

However, there is now compelling evidence that the active starburst was
not the only major starburst event to have occurred in M82.  A region at
$\sim 400-1000$ pc from the centre, ``M82 B'' (nomenclature from
O'Connell \& Mangano 1978), has the high surface brightness and spectral
features expected for an ancient starburst with an age in excess of
several 100 Myr and an amplitude similar to the active burst (O'Connell
\& Mangano 1978, Marcum \& O'Connell 1996, de Grijs et al.  2001,
hereafter dGOG).  Its spectral features resemble the characteristics of
the anomalous ``E+A'' spectra found in distant galaxy clusters, which
are generally interpreted as the signature of a truncated burst of star
formation that occurred $100-1000$ Myr earlier (e.g., Couch \& Sharples
1987, Dressler \& Gunn 1990, Couch et al.  1998).  It is thought that
these starbursts are closely related to the process by which disc
galaxies are converted into elliptical or lenticular galaxies and they
are thought to result from tidal interactions, mergers, or perhaps
ram-pressure stripping by the intergalactic medium (Butcher \& Oemler
1978, Oemler 1992, Barger et al.  1996). 

The significance of the detailed study of M82's starburst environment
lies, therefore, in the broader context of galaxy evolution.  Starbursts
of this scale are likely to be common features of early galaxy
evolution, and M82 is the nearest analogue to the sample of star-forming
galaxies recently identified at high redshifts ($z \gtrsim 3$; Steidel
et al.  1996, Giavalisco et al.  1997, Lowenthal et al.  1997).  In
addition, M82 affords a close-up view not only of an active starburst --
in M82 A, C and E -- but also, in region B and elsewhere, of the
multiple post-burst phases. 

In a recent study focusing on the fossil starburst site, M82 B, we found
a large population of ($\sim 110$) evolved compact star clusters (dGOG),
whose properties appear to be consistent with them being evolved (and
therefore faded) counterparts of the young (super) star clusters
detected in the galaxy's active core (O'Connell et al.  1995).  Based on
broad-band optical and near-infrared colours and comparison with stellar
evolutionary synthesis models, we estimated ages for the M82 B cluster
population from $\sim 30$ Myr to over 10 Gyr, with a peak near 650 Myr
(see also de Grijs 2001).  About 22 per cent of the clusters in M82 B
are older than 2 Gyr, with a roughly flat distribution to over 10 Gyr. 
Very few clusters are younger than 300 Myr.  The full-width of the peak
at 650 Myr is $\sim 500$ Myr, but this is undoubtedly broadened by the
various uncertainties entering the age-dating process.  The selection
bias of the star clusters in dGOG is such that the truncation of cluster
formation for ages $< 300$ Myr is better established than the roughly
constant formation rate at ages $> 2$ Gyr. 

Thus, we suggested steady, continuing cluster formation in M82 B at a
very modest rate at early times ($> 2$ Gyr ago) followed by a
concentrated formation episode lasting from 400--1000 Myr ago and a
subsequent suppression of cluster formation (dGOG). 

\subsection{Cluster disruption time-scales}

In the determination of the cluster formation history from the age
distribution of magnitude-limited cluster samples, cluster disruption
must be taken into account.  This is because the observed age
distribution is that of the surviving clusters only.  Therefore, in this
study of the evolution of the star cluster system in M82's fossil
starburst region, we determine both its cluster formation history and
the characteristic disruption time-scales. 

The dynamical evolution of star clusters is determined by a combination
of internal and external time-scales.  The free-fall and two-body
relaxation time-scales, which depend explicitly on the initial cluster
mass density (e.g., Spitzer 1957, Chernoff \& Weinberg 1990, de la
Fuente Marcos 1997, Portegies Zwart et al.  2001), affect the
cluster-internal processes of star formation and mass redistribution
through energy equipartition, leading to mass segregation and,
eventually, core collapse (see, e.g., de Grijs et al.  [2002a,b] for a
detailed description of mass segregation effects in young star
clusters).  While the internal relaxation process will, over time, eject
both high-mass stars from the core (e.g., due to interactions with hard
binaries; see Brandl et al.  2001, de Grijs et al.  2002a) and lose
lower-mass stars from its halo through diffusion (e.g., due to
Roche-lobe overflow), the external process of tidal disruption and
stripping by the surrounding galactic field is in general more important
for the discussion of the disruption of star clusters. 

Tidal disruptive processes are enhanced by ``normal'' stellar
evolutionary effects such as mass loss by winds and/or supernova
explosions, which will further reduce the stellar density in a cluster,
and thus make it more sensitive to external tidal forces. 

From the bimodal age distribution of (young) open and (old) globular
clusters in the Milky Way, Oort (1957) concluded that disruption of
Galactic star clusters must occur on time-scales of $\sim 5 \times 10^8$
yr.  Around the same time, Spitzer (1957) derived an expression for the
disruption time scale as a function of a cluster's mean density,
$\rho_c$ ($M_\odot$ pc$^{-3}$): $t_{\rm dis} = 1.9 \times 10^8 \rho_c$
yr, for $2.2 < \rho_c < 22 M_\odot$ pc$^{-3}$.  More advanced recent
studies, based on {\it N}-body modeling, have shown that the cluster
disruption time-scale is sensitive to the cluster mass, the fraction of
binary (or multiple) stars, and the initial mass function (IMF) adopted
(e.g., Chernoff \& Weinberg 1990, de la Fuente Marcos 1997). 

Boutloukos \& Lamers (2001, 2002) derived an empirical relation between
the disruption time and a cluster's initial mass for the Milky Way, the
Small Magellanic Cloud (SMC), M33 and the inner spiral arms of M51. 
They showed, based on an analysis of the mass and age distributions of
magnitude-limited samples of clusters, that the empirical disruption
time of clusters depends on their initial mass (\Mi) as
\begin{equation}
\tdis = \tdisref (\Mi/10^4 \Msun)^{\gamma}
\label{eq:tdis}
\end{equation}
where \tdisref\ is the disruption time of a cluster of initial mass
$\Mi=10^4 \Msun$.  The value of $\gamma$ is approximately the same in
these four galaxies, $\gamma = 0.62 \pm 0.06$.  However, the
characteristic disruption time-scale \tdisref\ is widely different in
the different galaxies.  The disruption time-scale is longest in the SMC
($\sim 8$ Gyr) and shortest in the inner spiral arms of M51 ($\sim 40$
Myr).  We will derive the disruption time-scale of clusters in M82 B and
allow for disruption in the determination of the cluster formation
history. 

In Section \ref{obs.sec}, we will summarise the observations on which
our discussion is based, and in Section \ref{theory.sec} we obtain new
age and mass estimates for the clusters in M82 B, based on the 3/2DEF
method (3/2-dimensional energy-fitting method; Bik et al.  2002) of
fitting the observed spectral energy distribution to that of cluster
evolution models.  In Section \ref{history.sec} we derive the cluster
formation history and the characteristic cluster disruption time-scale. 
In Section \ref{sec:burstmass} we discuss the mass distribution of the
clusters formed during the burst.  The methods and assumptions are
discussed in Section \ref{discussion.sec}; finally, we will summarise
and conclude the paper in Section \ref{summary.sec}.

\section{Observations}
\label{obs.sec}

The observations on which our analysis of the M82 cluster formation
history is based were described in detail in dGOG.  Briefly summarised,
we observed region B with the {\sl Wide Field Planetary Camera 2
(WFPC2)} on board the {\sl Hubble Space Telescope (HST)} through the
F439W, F555W and F814W filters, roughly corresponding to the standard
Johnson-Cousins {\it B, V} and {\it I} passbands, respectively.  We
obtained {\sl WFPC2} observations using two pointings, so that the
entire region B was covered by the Planetary Camera (PC) chip, the
highest-resolution optical imaging instrument available on board {\sl
HST} at that time, with a pixel size of $0.0455''$.  The effective
integration times used for the F439W, F555W, and F814W observations were
4400s, 2500s and 2200s, respectively, for the western half of M82 B and
4100s, 3100s and 2200s, respectively, for the eastern half closest to
the galaxy's starburst core. 

In addition, we imaged the entire region with {\sl HST}'s {\sl
Near-Infrared Camera and Multi-Object Spectrometer (NICMOS)} (Camera 2;
pixel size $0.075''$) in both the F110W and F160W filters (comparable to
the Bessell {\it J} and {\it H} filters, respectively), in a tiled
pattern of $2 \times 4$ partially overlapping exposures.  The
integrations, with effective integration times of 768s for each of the
eastern and western halfs of M82 B and each filter, were taken in
MULTIACCUM mode to preserve dynamic range and to correct for cosmic
rays. 

We subsequently obtained integrated photometry for the extended objects
(i.e., star clusters) in M82 B down to a 50 per cent completeness limit
of $V \simeq 23.3$ mag.  Because of the highly variable background and
the numerous spurious features due to dust lanes and background
variations, we decided to select only genuine star cluster brighter than
$V = 22.5$ for further analysis, corresponding to close to 100 per cent
completeness (dGOG). 

\section{Observed cluster mass and age distributions}
\label{theory.sec}

For the final 113 objects obtained following the procedures outlined in
the previous section, dGOG used $(B-V)$ versus $(V-I)$ colour-colour
diagrams to disentangle age and extinction effects, since the age and
extinction vectors are not entirely degenerate for this choice of
optical colours. 

Accurate age and mass determinations are essential for the analysis of
cluster formation and disruption time-scales performed in this paper. 
Combining the observed luminosities of the M82 B star cluster population
with the appropriate, age-dependent mass-to-light ratios from spectral
synthesis models provided us with photometric mass estimates with an
accuracy of well within an order of magnitude.  Independent dynamical
mass estimates from high-resolution spectroscopy are available only for
a few of the most luminous super star clusters seen in the nearest
starburst galaxies (M82, the Antennae, NGC 1569 and NGC 1705), and are
approximately $10^6 M_\odot$ (Ho \& Filippenko 1996a,b, Smith \&
Gallagher 2000, Mengel et al.  2002).  Because of the proximity of M82,
we have been able to probe the young cluster population in M82 B to
fainter absolute magnitudes, and thus lower masses, than has been
possible before in other starburst galaxies.  Other young star cluster
samples are biased towards high masses by selection effects due to their
host galaxies' greater distances. 

\begin{figure*}
\hspace{1.2cm}
\psfig{figure=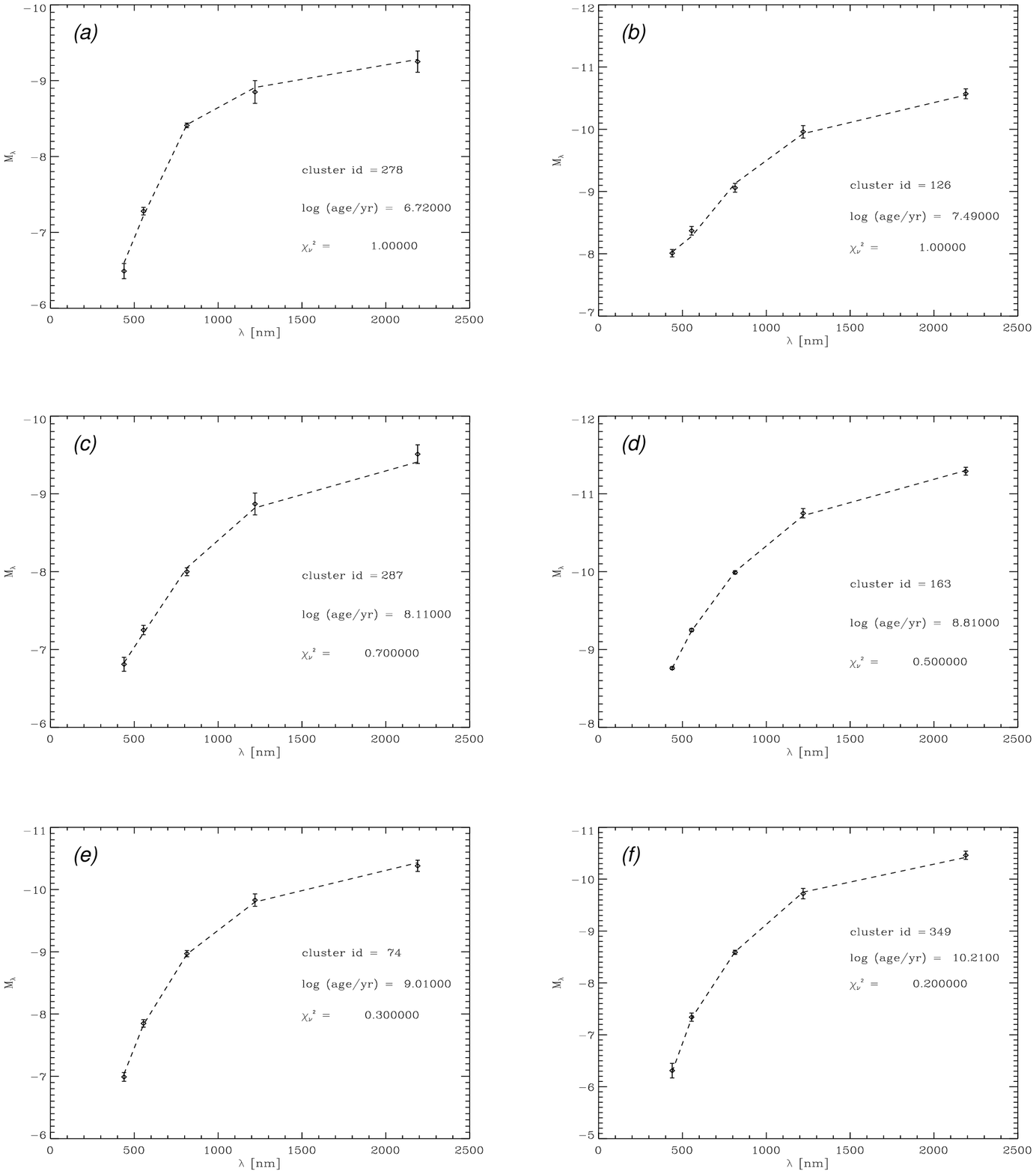,width=15cm}
\caption{\label{composite.fig}Examples of a few fits of the Starburst99
and BC00 models (dashed lines) to the observed cluster SEDs sampled in
the $B, V, I, J,$ and $H$ passbands (data points with error bars),
representative of clusters throughout the entire age range (age
increases from panel a to f).  The best-fitting model SEDs are based on
the 3/2DEF maximum likelihood method.}
\end{figure*}

Keeping the importance of the age dependence on our photometric mass
estimates in mind, we decided to redetermine the M82 B cluster ages
using the full parameter space available.  We obtained more accurate age
determinations by also including the NICMOS observations, and solving
for the best-fitting ages matching the full spectral energy
distributions (SEDs) of our sample clusters, from the F439W to the F160W
passband.  We compared the observed cluster SEDs with the model
predictions for an instantaneous burst of star formation, assuming a
Salpeter IMF from $0.1-100 M_\odot$ with power-law slope $\alpha_{\rm
IMF} = 2.35$ to obtain our new estimates for the cluster age {\it t},
initial mass $M_{\rm cl}$ and extinction $E(B-V)$; for the latter we
adopted the (Galactic) extinction law of Scuderi et al.  (1996).  For
ages $t \le 10^9$ yr we used the Starburst99 models (Leitherer et al. 
1999), while for older ages we used the most recent single stellar
population models by Bruzual \& Charlot (2000, hereafter BC00). 
Examples of our model fits to the observed cluster SEDs are shown in
Fig.  \ref{composite.fig}. 

We realise that recent determinations of the stellar IMF deviate
significantly from a Salpeter-type IMF at low masses, in the sense that
the low-mass stellar IMF is significantly flatter than the Salpeter
slope.  The implication of using a Salpeter-type IMF for our cluster
mass determinations is therefore that we have {\it overestimated} the
individual cluster masses (although the relative mass distribution of
our entire cluster sample remains unaffected).  Therefore, we used the
more modern IMF of Kroupa, Tout \& Gilmore (1993, hereafter KTG) to
determine the correction factor, $C$, between our masses and the more
realistic masses obtained from the KTG IMF (both normalised at $1.0
M_\odot$).  This IMF is characterised by slopes of $\alpha = -2.7$ for
$m > 1.0 M_\odot$, $\alpha = -2.2$ for $0.5 \le m/M_\odot \le 1.0$, and
$-1.85 < \alpha < -0.70$ for $0.08 < m/M_\odot \le 0.5$.  Depending on
the adopted slope for the lowest mass range, we have therefore
overestimated our individual cluster masses by a factor of $1.70 < C <
3.46$ for an IMF containing stellar masses in the range $0.1 \le
m/M_\odot \le 100$. 

The fitting of the observed cluster SEDs to the Starburst99 and BC00
models was done using a three-dimensional maximum likelihood method,
3/2DEF, with the initial mass $\Mi$, age and extinction $E(B-V)$ as free
parameters (see Bik et al.  2002).  For the 46 clusters with upper
limits in one or more filters we used a two-dimensional maximum
likelihood fit, using the extinction probability distribution for
$E(B-V)$.  This distribution was derived for the clusters with
well-defined SEDs over the full wavelength range (see Bik et al.  2002). 
We obtained reliable age estimates based on the full SED modeling for 81
of the 113 clusters.  The new age and mass estimates for these 81
clusters are listed in Table \ref{agemass.tab}, where the individual
clusters are identified by their dGOG ID.  The dGOG mass estimates in
Table \ref{agemass.tab} were determined by combining the absolute
magnitudes of the clusters with their age-dependent mass-to-light
ratios. 

\begin{table*}
\caption[ ]{\label{agemass.tab}New age, mass and extinction
determinations for the M82 B clusters, and comparison with previously
published values. 
}
{\scriptsize
\begin{center}
\begin{tabular}{crrrcccccccrcc}
\hline
& \multicolumn{9}{c}{This paper} & & \multicolumn{3}{c}{dGOG} \\
\cline{2-10}\cline{12-14}
\multicolumn{1}{c}{Object} & \multicolumn{3}{c}{log(Age/yr)} &
\multicolumn{3}{c}{log($m/M_\odot$)$^a$} & \multicolumn{3}{c}{E$(B-V)$
(mag)} & & \multicolumn{1}{c}{log(Age/yr)} &
\multicolumn{1}{c}{log($m/M_\odot$)} & \multicolumn{1}{c}{E$(B-V)$} \\
& \multicolumn{1}{c}{min} & \multicolumn{1}{c}{best} &
\multicolumn{1}{c}{max} & \multicolumn{1}{c}{min} &
\multicolumn{1}{c}{best} & \multicolumn{1}{c}{max} &
\multicolumn{1}{c}{min} & \multicolumn{1}{c}{best} &
\multicolumn{1}{c}{max} & & & & \multicolumn{1}{c}{(mag)$^b$} \\
\hline
B1-01 &  9.16 &  9.88 & 10.05 & 5.20 & 5.62 & 5.74 & 0.00 & 0.02 & 0.20 & & 10.24 & 5.92 & 0.31 \\
B1-02 &  8.96 &  9.28 &  9.36 & 4.66 & 4.90 & 4.98 & 0.00 & 0.00 & 0.16 & &  9.27 & 4.90 & 0.04 \\
B1-04 &  5.10 &  8.46 &  8.81 & 4.14 & 5.21 & 5.34 & 0.06 & 0.18 & 0.70 & &  8.78 & 5.25 & 0.58 \\
B1-05 &  9.74 & 10.05 & 10.15 & 5.76 & 5.95 & 6.03 & 0.00 & 0.00 & 0.06 & & 10.14 & 6.04 & 0.18 \\
B1-06 &  5.10 &  6.58 &  8.76 & 4.33 & 4.53 & 5.54 & 0.00 & 0.52 & 0.60 & &  8.67 & 5.46 & 0.44 \\
B1-07 &  8.96 &  9.01 &  9.16 & 5.53 & 5.60 & 5.69 & 0.00 & 0.06 & 0.08 & &  9.03 & 5.53 & 0.12 \\
B1-08 &  8.51 &  8.66 &  8.66 & 5.14 & 5.24 & 5.24 & 0.00 & 0.02 & 0.04 & &  8.68 & 5.24 & 0.09 \\
B1-09 &  5.10 &  8.91 &  9.01 & 3.14 & 4.24 & 4.34 & 0.00 & 0.00 & 0.72 & &  8.89 & 4.21 & 0.00 \\
B1-10 &  5.10 &  8.71 &  8.86 & 3.14 & 4.33 & 4.41 & 0.00 & 0.04 & 0.64 & &  8.76 & 4.33 & 0.18 \\
B1-11 &  8.71 &  8.86 &  8.86 & 5.36 & 5.38 & 5.41 & 0.00 & 0.00 & 0.08 & &  8.84 & 5.36 & 0.09 \\
B1-12 &  8.71 &  9.01 &  9.21 & 4.81 & 4.88 & 5.02 & 0.00 & 0.06 & 0.22 & &  9.03 & 4.81 & 0.06 \\
B1-14 &  8.51 &  8.86 &  9.01 & 5.11 & 5.24 & 5.28 & 0.00 & 0.10 & 0.22 & &  8.89 & 5.13 & 0.34 \\
B1-16 &  8.56 &  9.95 & 10.30 & 4.94 & 5.63 & 5.87 & 0.00 & 0.04 & 0.50 & & 10.00 & 5.63 & 0.40 \\
B1-17 &  8.66 &  9.01 &  9.16 & 4.47 & 4.56 & 4.68 & 0.00 & 0.04 & 0.24 & &  9.03 & 4.50 & 0.17 \\
B1-18 &  8.86 &  8.91 &  8.91 & 5.54 & 5.56 & 5.56 & 0.00 & 0.00 & 0.02 & &  8.85 & 5.50 & 0.15 \\
B1-20 &  9.01 &  9.16 &  9.16 & 5.48 & 5.65 & 5.65 & 0.00 & 0.00 & 0.08 & &  9.10 & 5.60 & 0.07 \\
B1-22 &  8.56 &  8.76 &  8.86 & 5.23 & 5.29 & 5.33 & 0.00 & 0.04 & 0.12 & &  8.78 & 5.26 & 0.25 \\
B1-24 &  8.66 &  8.86 &  8.91 & 5.15 & 5.20 & 5.23 & 0.00 & 0.02 & 0.12 & &  8.83 & 5.15 & 0.13 \\
B1-25 &  8.91 &  9.01 &  9.26 & 5.00 & 5.10 & 5.23 & 0.00 & 0.10 & 0.16 & &  9.05 & 4.97 & 0.16 \\
B1-26 &  5.10 &  8.61 &  8.66 & 3.09 & 4.26 & 4.32 & 0.00 & 0.00 & 0.56 & &  8.68 & 4.33 & 0.00 \\
B1-27 &  8.76 &  8.86 &  8.91 & 4.44 & 4.49 & 4.51 & 0.00 & 0.02 & 0.08 & &  8.86 & 4.46 & 0.00 \\
B1-28 &  8.86 &  8.86 &  8.86 & 5.97 & 5.97 & 5.97 & 0.04 & 0.04 & 0.04 & &  8.86 & 5.90 & 0.13 \\
B1-29 &  5.10 &  8.76 &  8.91 & 3.52 & 4.71 & 4.78 & 0.00 & 0.06 & 0.68 & &  8.81 & 4.68 & 0.25 \\
B1-32 &  8.76 &  9.01 &  9.01 & 4.64 & 4.69 & 4.72 & 0.00 & 0.02 & 0.14 & &  8.98 & 4.65 & 0.10 \\
B1-34 &  5.10 &  6.84 &  8.36 & 3.31 & 3.39 & 4.44 & 0.00 & 0.12 & 0.46 & &  7.55 & 3.94 & 0.35 \\
B1-35 &  8.61 &  8.86 &  9.01 & 5.02 & 5.08 & 5.12 & 0.00 & 0.08 & 0.20 & &  8.88 & 4.98 & 0.24 \\
B1-36 &  8.96 & 10.14 & 10.30 & 5.13 & 5.90 & 6.01 & 0.00 & 0.00 & 0.34 & & 10.14 & 5.91 & 0.24 \\
B1-37 &  9.16 &  9.40 &  9.44 & 5.75 & 5.88 & 5.91 & 0.00 & 0.00 & 0.10 & &  9.32 & 5.80 & 0.18 \\
B1-38 &  5.10 &  8.81 &  8.96 & 3.57 & 4.63 & 4.71 & 0.00 & 0.04 & 0.68 & &  8.83 & 4.60 & 0.16 \\
B1-39 &  8.81 &  9.34 &  9.44 & 4.29 & 4.58 & 4.70 & 0.00 & 0.00 & 0.28 & &  9.28 & 4.53 & 0.15 \\
B1-40 &  5.10 &  8.66 &  8.76 & 2.79 & 3.97 & 4.03 & 0.00 & 0.02 & 0.60 & &  8.68 & 3.97 & 0.13 \\
B1-43 &  9.01 &  9.48 &  9.68 & 4.35 & 4.69 & 4.87 & 0.00 & 0.00 & 0.20 & &  9.70 & 4.90 & 0.11 \\
B2-01 &  9.01 &  9.94 & 10.25 & 4.43 & 5.09 & 5.34 & 0.00 & 0.00 & 0.28 & & 10.00 & 5.16 & 0.00 \\
B2-04 &  9.01 &  9.23 &  9.28 & 5.23 & 5.44 & 5.47 & 0.00 & 0.02 & 0.14 & &  9.32 & 5.51 & 0.15 \\
B2-05 &  5.10 &  8.46 &  8.71 & 4.14 & 5.21 & 5.35 & 0.00 & 0.08 & 0.62 & &  8.66 & 5.28 & 0.32 \\
B2-07 &  8.66 &  8.96 &  9.41 & 5.41 & 5.55 & 5.78 & 0.00 & 0.18 & 0.34 & &  9.19 & 5.56 & 0.43 \\
B2-08 &  8.66 &  9.01 & 10.26 & 5.49 & 5.60 & 6.35 & 0.00 & 0.32 & 0.50 & &  9.72 & 5.89 & 0.77 \\
B2-12 &  9.74 &  9.76 &  9.80 & 6.86 & 6.88 & 6.91 & 0.00 & 0.00 & 0.00 & &  9.70 & 6.83 & 0.22 \\
B2-13 &  9.30 &  9.83 &  9.90 & 5.38 & 5.71 & 5.77 & 0.00 & 0.00 & 0.12 & &  9.76 & 5.66 & 0.19 \\
B2-14 &  8.71 &  9.44 & 10.14 & 4.75 & 5.13 & 5.63 & 0.00 & 0.08 & 0.40 & & 10.13 & 5.64 & 0.43 \\
B2-17 &  8.96 &  9.48 &  9.80 & 5.21 & 5.56 & 5.81 & 0.00 & 0.02 & 0.26 & &  9.46 & 5.52 & 0.26 \\
B2-18 &  5.10 &  7.49 &  8.66 & 4.09 & 4.79 & 5.44 & 0.08 & 0.26 & 0.74 & &  9.23 & 5.69 & 1.03 \\
B2-21 &  6.86 &  7.00 &  7.60 & 4.46 & 4.57 & 5.17 & 0.00 & 0.04 & 0.24 & &  8.90 & 5.82 & 1.13 \\
B2-22 &  8.96 & 10.20 & 10.30 & 4.53 & 5.36 & 5.47 & 0.00 & 0.00 & 0.36 & & 10.26 & 5.39 & 0.15 \\
B2-25 &  5.10 &  8.51 &  8.81 & 3.90 & 5.01 & 5.14 & 0.00 & 0.10 & 0.64 & &  8.71 & 5.05 & 0.29 \\
B2-26 &  8.71 &  8.76 &  8.76 & 5.60 & 5.62 & 5.62 & 0.02 & 0.02 & 0.04 & &  8.77 & 5.63 & 0.14 \\
B2-28 &  5.10 &  7.49 &  8.36 & 3.51 & 4.09 & 4.65 & 0.00 & 0.00 & 0.48 & &  7.26 & 3.91 & 0.60 \\
B2-29 &  5.10 &  6.66 &  8.71 & 2.95 & 3.21 & 4.22 & 0.00 & 0.44 & 0.58 & &  8.71 & 4.21 & 0.25 \\
B2-30 &  8.41 &  8.71 &  9.01 & 4.99 & 5.12 & 5.19 & 0.00 & 0.16 & 0.24 & &  8.89 & 5.05 & 0.37 \\
B2-31 &  5.10 &  8.51 &  8.71 & 3.21 & 4.29 & 4.39 & 0.00 & 0.06 & 0.60 & &  8.65 & 4.34 & 0.24 \\
B2-32 &  8.81 &  9.16 &  9.60 & 5.05 & 5.23 & 5.50 & 0.00 & 0.12 & 0.28 & &  9.83 & 5.70 & 0.37 \\
B2-33 &  5.10 &  8.71 &  9.01 & 3.03 & 4.27 & 4.42 & 0.00 & 0.12 & 0.78 & &  8.83 & 4.23 & 0.24 \\
B2-34 &  5.10 &  6.48 &  8.71 & 2.97 & 3.32 & 4.19 & 0.00 & 0.54 & 0.58 & &  8.56 & 4.08 & 0.29 \\
B2-36 &  9.16 &  9.16 &  9.21 & 5.32 & 5.34 & 5.36 & 0.00 & 0.02 & 0.02 & &  9.20 & 5.38 & 0.00 \\
B2-37 &  6.84 &  7.14 &  7.65 & 4.42 & 4.70 & 5.11 & 0.00 & 0.00 & 0.16 & &  7.51 & 4.97 & 0.69 \\
B2-38 &  5.10 &  8.76 &  8.96 & 3.08 & 4.34 & 4.52 & 0.00 & 0.00 & 0.70 & &  8.83 & 4.42 & 0.00 \\
B2-39 &  8.71 &  8.76 &  8.81 & 4.56 & 4.57 & 4.62 & 0.00 & 0.00 & 0.04 & &  8.83 & 4.65 & 0.00 \\
B2-40 &  8.41 &  8.46 &  8.56 & 5.07 & 5.08 & 5.14 & 0.00 & 0.02 & 0.04 & &  8.51 & 5.10 & 0.13 \\
B2-41 &  8.96 &  9.01 &  9.01 & 5.91 & 5.92 & 5.92 & 0.04 & 0.04 & 0.06 & &  9.00 & 5.86 & 0.20 \\
B2-43 &  8.66 &  8.86 &  8.91 & 5.15 & 5.20 & 5.23 & 0.00 & 0.02 & 0.12 & &  8.81 & 4.54 & 0.00 \\
B2-44 &  5.10 &  8.66 &  8.76 & 3.57 & 4.61 & 4.64 & 0.00 & 0.04 & 0.60 & &  8.70 & 4.61 & 0.12 \\
B2-45 &  8.71 &  9.01 &  9.70 & 5.00 & 5.09 & 5.51 & 0.00 & 0.20 & 0.36 & &  9.93 & 5.70 & 0.37 \\
B2-47 &  9.01 &  9.26 &  9.32 & 4.12 & 4.36 & 4.42 & 0.00 & 0.00 & 0.12 & &  9.28 & 4.39 & 0.00 \\
B2-48 &  5.10 &  8.61 &  8.81 & 3.32 & 4.52 & 4.61 & 0.00 & 0.06 & 0.64 & &  8.68 & 4.50 & 0.24 \\
B2-50 &  5.10 &  7.68 &  8.51 & 3.54 & 4.24 & 4.69 & 0.00 & 0.06 & 0.52 & &  8.31 & 4.57 & 0.48 \\
B2-51 &  8.96 & 10.14 & 10.30 & 5.13 & 5.90 & 6.01 & 0.00 & 0.00 & 0.34 & &  8.63 & 3.90 & 0.00 \\
B2-52 &  8.96 &  9.01 &  9.01 & 4.66 & 4.67 & 4.71 & 0.00 & 0.00 & 0.06 & &  9.01 & 4.67 & 0.00 \\
B2-54 &  5.10 &  8.66 &  8.71 & 3.39 & 4.44 & 4.47 & 0.00 & 0.02 & 0.58 & &  8.68 & 4.44 & 0.00 \\
B2-55 &  8.56 &  8.71 &  8.76 & 3.87 & 3.94 & 3.98 & 0.00 & 0.00 & 0.04 & &  8.78 & 4.03 & 0.00 \\
B2-56 &  8.86 &  9.26 &  9.36 & 4.76 & 5.00 & 5.10 & 0.00 & 0.00 & 0.22 & &  9.23 & 4.98 & 0.00 \\
B2-57 &  8.86 &  9.23 &  9.41 & 5.21 & 5.42 & 5.55 & 0.00 & 0.04 & 0.22 & &  9.26 & 5.40 & 0.26 \\
B2-59 &  5.10 &  8.21 &  8.96 & 3.69 & 4.73 & 5.00 & 0.00 & 0.24 & 0.72 & &  8.75 & 4.82 & 0.65 \\
B2-60 &  5.10 &  8.76 &  8.86 & 3.46 & 4.48 & 4.54 & 0.00 & 0.02 & 0.64 & &  8.78 & 4.49 & 0.18 \\
B2-62 &  8.76 &  8.96 &  9.01 & 4.05 & 4.12 & 4.14 & 0.00 & 0.02 & 0.12 & &  8.95 & 4.09 & 0.00 \\
B2-64 &  8.96 &  9.23 &  9.30 & 4.25 & 4.48 & 4.55 & 0.00 & 0.00 & 0.14 & &  9.26 & 4.52 & 0.00 \\
B2-65 &  8.91 &  9.11 &  9.16 & 5.02 & 5.12 & 5.20 & 0.00 & 0.00 & 0.12 & &  9.03 & 5.01 & 0.20 \\
B2-67 &  8.81 &  8.91 &  8.96 & 5.22 & 5.25 & 5.27 & 0.06 & 0.08 & 0.12 & &  8.90 & 5.11 & 0.35 \\
B2-68 &  8.86 &  9.16 &  9.30 & 4.51 & 4.69 & 4.77 & 0.00 & 0.04 & 0.18 & &  9.20 & 4.70 & 0.07 \\
B2-69 &  8.86 &  8.96 &  8.96 & 4.09 & 4.14 & 4.16 & 0.00 & 0.00 & 0.04 & &  8.85 & 4.04 & 0.00 \\
B2-70 &  8.76 &  8.81 &  8.81 & 4.97 & 4.99 & 4.99 & 0.02 & 0.02 & 0.04 & &  8.81 & 4.97 & 0.03 \\
\hline
\end{tabular}
\end{center}
\flushleft
{\sc Notes:} $^a$ based on a Salpeter-type IMF; comparison with results
from more modern IMFs (e.g., KTG) implies that we have overestimated our
individual cluster masses by a factor of about 1.7--3.5, depending on
the adopted IMF slope for the lowest stellar masses (see Sect. 
\ref{theory.sec}); $^b$ E$(B-V) = A_V / 3.1$. 
}
\end{table*}

Our analysis in both dGOG and the present paper is based on the
assumption of solar metallicity, which should be a reasonable match to
the young objects in M82 (Gallagher \& Smith 1999; see also Fritze-v. 
Alvensleben \& Gerhard 1994), but the effects of varying the metallicity
between one-fifth and 2.5 times solar are small compared to the
photometric uncertainties (see also dGOG).  We therefore assert that our
results are not significantly affected by possibly varying
metallicities.  Ongoing analysis (Parmentier, de Grijs \& Gilmore, 2002)
suggests that this assumption was indeed justified.  We should caution,
however, that we have only been able to sample the star clusters located
close to the surface of region B, as evidenced by the derived extinction
estimates: dGOG find $A_V \lesssim 1$ mag in general, while our new
determinations (this paper) restrict the extinction even more, to $A_V
\lesssim 0.2$ mag for the clusters with well-determined ages, and
possibly up to $A_V \sim 0.40-0.55$ for some of the others. 

\begin{figure*}
\psfig{figure=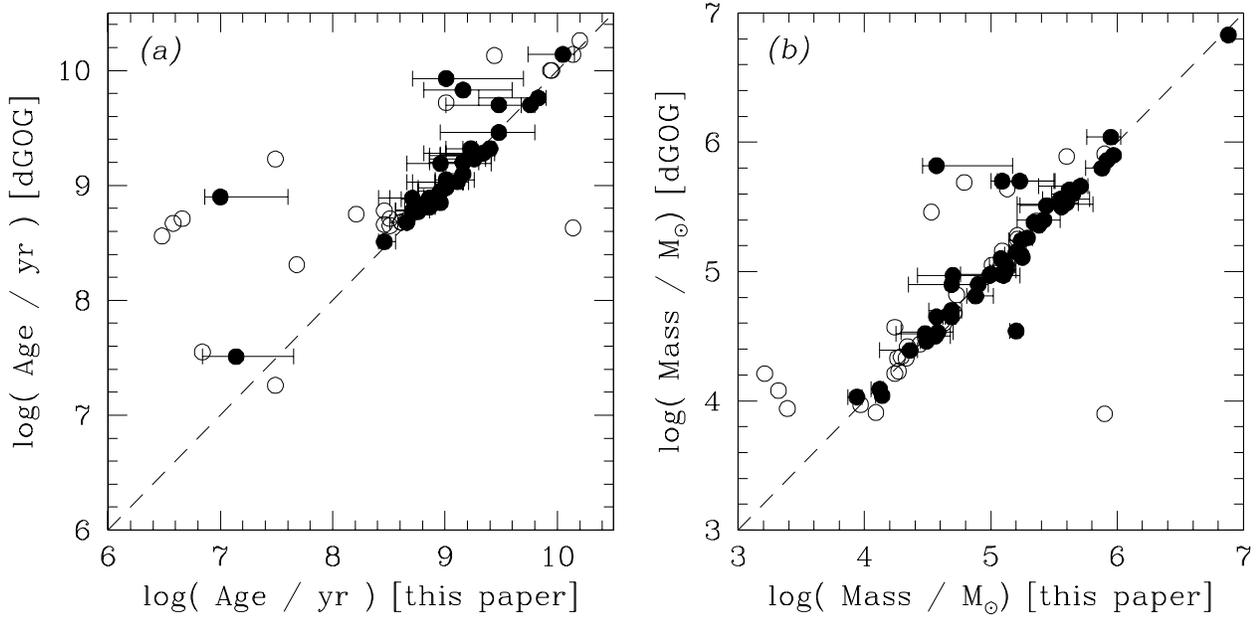,width=18cm}
\vspace*{-8.5cm}
\caption{\label{compare.fig}Comparison between {\it (a)} the age and
{\it (b)} mass estimates of dGOG based on {\it BVI} photometry and new
estimates using additional {\it JH} observations and a full-parameter
fitting routine as described in the text.  The black dots represent
clusters for which the total age range obtained is ($\log( {\rm
Age[max]} ) - \log( {\rm Age[min]} )) \le 1.0$; the open circles are
objects with more uncertain age determinations. They show a larger
scatter than the black dots, but both distributions are internally
consistent.}
\end{figure*}

In Fig.  \ref{compare.fig} we compare the age and mass estimates from
the more sophisticated 3/2DEF method used in this paper with the
corresponding parameters obtained by dGOG based on their location in the
$(B-V)$ vs.  $(V-I)$ diagrams.  The black dots represent clusters for
which the range between the minimum and maximum plausible ages, $\log(
{\rm Age[max]} ) - \log( {\rm Age[min]} ) \le 1.0$; the open circles are
objects with more uncertain age determinations.  We conclude that the
new age and mass determinations are consistent with the dGOG values;
this consistency is better for the photometric cluster mass estimates
than for their ages, as shown by the larger scatter in Fig. 
\ref{compare.fig}a with respect to panel b.  However, we also note that
the age and mass distributions obtained for the subsample with
well-determined ages and for the full sample are internally consistent. 

\begin{figure*}
\hspace*{1.5cm}
\psfig{figure=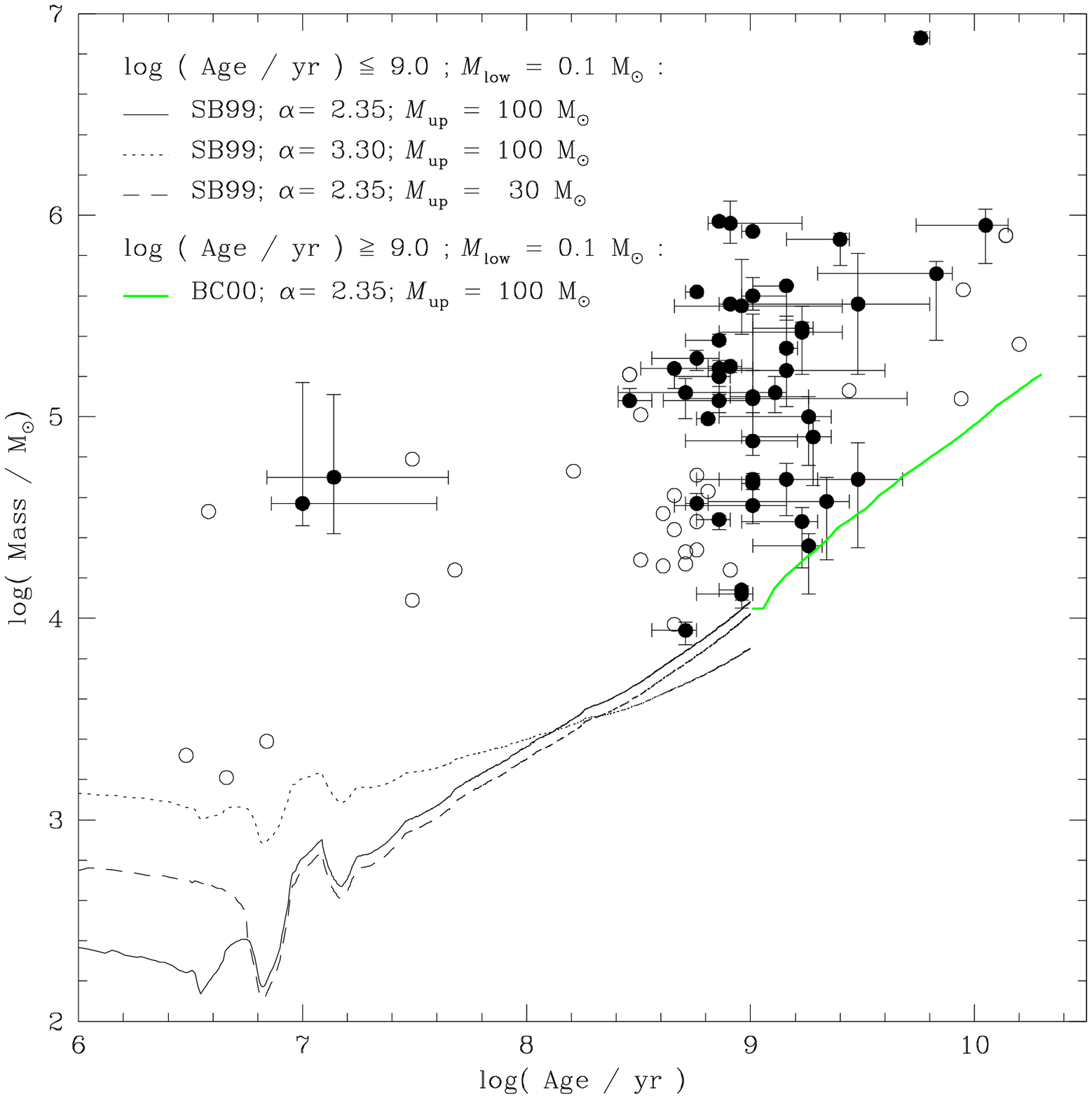,width=15cm}
\vspace*{-0.5cm}
\caption{\label{agemass.fig}Distribution of the M82 B clusters in the
(age vs.  mass) plane.  The symbol coding is as in Fig. 
\ref{compare.fig}.  Overplotted for ages up to 1.0 Gyr are the expected
detection limits in the (age vs.  mass) plane, predicted by Starburst99
for a range of IMFs; for older ages, we use the BC00 models for a
standard IMF, as indicated in the figure legend.  These model
predictions are based on a detection limit of $V = 22.5$ and
$(m-M)_{{\rm M}82} = 27.8$, assuming no extinction.  For a nominal
extinction of $A_V = 0.2$ mag, expected for the clusters with
well-determined ages, the detection limit is expected to shift to higher
masses by $\Delta \log( M_{\rm cl}/ M_\odot ) = 0.08$, which is well
within the uncertainties associated with our mass determinations.  The
features around 10 Myr are caused by the appearance of red supergiants.}
\end{figure*}

Fig.  \ref{agemass.fig} shows the distribution of the M82 B clusters in
the age vs.  mass plane.  It is immediately clear that the lower mass
limit increases with increasing cluster age.  The various (solid, dashed
and dotted) lines overplotted on the figure show that this is indeed the
expected effect of normal evolutionary fading of a synthetic single
stellar population of an instantaneously formed cluster.  For ages $\le
10^9$ yr, we show the unreddened fading line for clusters with a
limiting magnitude of $V = 22.5$ at the distance of M82 ($m-M = 27.8$;
see dGOG), and various choices for the IMF (indicated in the figure are
the IMF slope $\alpha_{\rm IMF}$, and the lower and upper mass
cut-offs), predicted by the Starburst99 models (``SB99'').  For older
ages ($t \ge 10^9$ yr), we show its extension predicted by the BC00
models.  These predicted lower limits agree well with our data points. 

\section{The derived cluster formation history}
\label{history.sec}

\subsection{Cluster age and mass distributions}
\label{distributions.sec}

\begin{figure*}
\hspace{1.2cm}
\psfig{figure=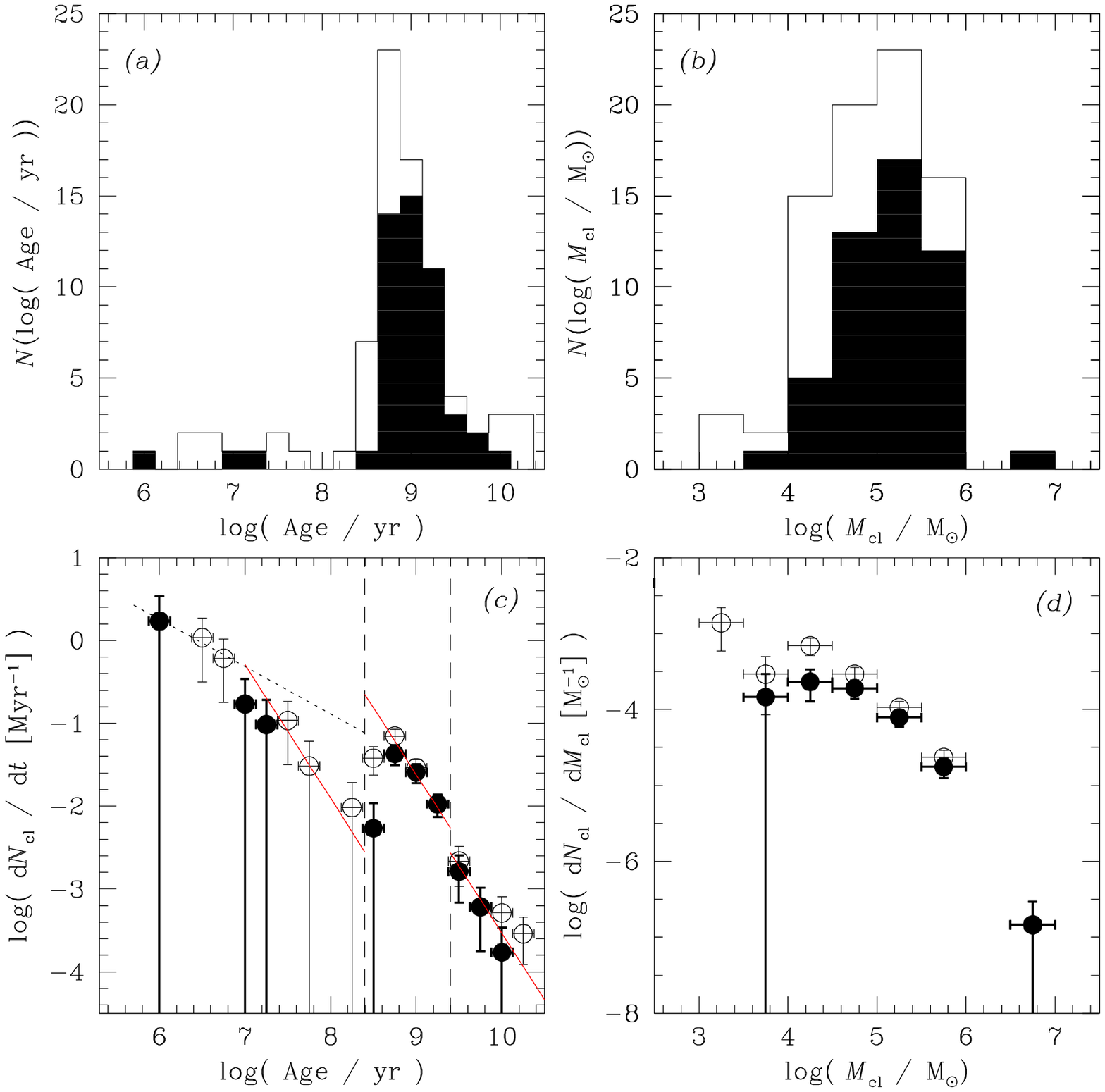,width=15cm}
\caption{\label{rates.fig}{\it (a)} and {\it (b)} -- Age and mass
distributions of the M82 B clusters, redetermined in this paper using
improved methods.  The shaded histograms correspond to the clusters with
($\log( {\rm Age[max]} ) - \log( {\rm Age[min]} ) \le 1.0$; the open
histograms represent the entire cluster sample.  {\it (c)} -- The
cluster formation rate (in number of clusters per Myr) as a function of
age.  Open circles: full sample; filled circles: clusters with
well-determined ages, as above.  The vertical dashed lines indicate the
age range dominated by the burst of cluster formation.  The dotted line
is the least-squares power-law fit to the fading, non-disrupted
clusters, for a constant ongoing cluster formation rate.  The solid line
segments are the disruption lines of clusters formed in the pre-burst
phase, during the burst, and in the post-burst phase, as described in
Section \ref{timescales.sec}.  {\it (d)} -- Mass spectrum of the M82 B
clusters (number of clusters per mass bin); symbol coding as in panel
{\it (c)}.}
\end{figure*}

The cluster age and mass distributions for the M82 B cluster sample are
shown in Figs.  \ref{rates.fig}a and b.  Our new age estimates confirm
the peak in the age histogram attributed to the last tidal encounter
with M81 by dGOG.  For the subsample of clusters with well-determined
ages (shaded histogram), we find a peak formation epoch at $\log( t_{\rm
peak} / {\rm yr} ) = 9.04 \; (t_{\rm peak} \simeq 1.10$ Gyr), with a
Gaussian $\sigma$ of $\Delta \log( t_{\rm width}) = 0.27$, corresponding
to a FWHM of $\Delta \log( t_{\rm width}) = 0.64$.  The corresponding
numbers for the full sample are $\log( t_{\rm peak} / {\rm yr} ) = 8.97
\; (t_{\rm peak} \simeq 0.93$ Gyr) for the peak of cluster formation,
and for $\sigma$, $\Delta \log( t_{\rm width}) = 0.35$ (FWHM, $\Delta
\log( t_{\rm width}) = 0.82$).  In dGOG (see also de Grijs 2001), we
concluded that there is a strong peak of cluster formation at $\sim 650$
Myr ago, which have formed over a period of $\sim 500$ Myr, but very few
clusters are younger than $\log( t / {\rm yr} ) \simeq 8.5$ ($t \simeq
300$ Myr).  Our new age estimates date the event triggering the
starburst to be slightly older than the value of 650 Myr derived by
dGOG.  Our estimate of the duration of the burst should be considered an
upper limit, because uncertainties in the cluster age determination may
have broadened the peak in Fig.  \ref{rates.fig}a. 

At the distance from the centre of region B, one would expect M82's
differential rotation (cf.  Shen \& Lo 1995) to have caused the
starburst area to disperse on these time-scales.  The reason why the
fossil starburst region has remained relatively well constrained is
likely found in the complex structure of the disc.  It is well-known
that the inner $\sim 1$ kpc of M82 is dominated by a stellar bar (e.g.,
Wills et al.  2000) in solid-body rotation.  From observations in other
galaxies, it appears to be a common feature that central bars are often
surrounded by a ring-like structure.  If this is also true for M82, it
is reasonable to assume that stars in the ring are trapped, and
therefore cannot move very much in radius because of dynamical resonance
effects.  The phase mixing around the ring might be slow enough for a
specific part of the ring to keep its identity over a sufficient time so
as to appear like region B (see de Grijs 2001): if the diffusion
velocity around the ring is sufficiently small, any specific region
would remain self-constrained for several rotation periods.  In
addition, since the density in the region is high (see Sect. 
\ref{timescales.sec}), simple calculations imply that the area's
self-gravity is non-negligible compared to the rotational shear,
therefore also prohibiting a rapid dispersion of the fossil starburst
region. 

The improved mass estimates also confirm that the (initial) masses of
the young clusters in M82 B with $V \le 22.5$ mag are mostly in the
range $10^4 - 10^6 M_\odot$, with a median of $10^5 M_\odot$ (dGOG). In
fact, based on a Gaussian fit, we find that the mean mass of our M82 B
cluster sample is $\log( M_{\rm cl} / M_\odot ) = 5.03$ and 4.88 for the
subsample with well-determined ages and the full sample, respectively,
corresponding to $M_{\rm cl} = 1.1 \times 10^5$ and $0.8 \times 10^5
M_\odot$, respectively.

\subsection{Cluster disruption and the cluster formation history}
\label{assumptions.sec}

In Figs.  \ref{rates.fig}c and d we show the formation rate and the mass
spectrum of the observed clusters in M82 B.  The open circles were
derived for the entire sample, whereas the filled data points represent
only those clusters for which we could obtain good age estimates.  These
distributions depend on the cluster formation history and on the cluster
disruption time-scale of M82 B. 

As shown by Boutloukos \& Lamers (2002, hereafter Paper I), with only a few
well-justified assumptions, the mass and age distributions
of a magnitude-limited sample of star clusters in a
given galaxy can be predicted both accurately and robustly, despite the
complex physical processes underlying the assumptions (for a full
discussion see Paper I). If all of the following conditions are met:

\begin{enumerate}
\item the cluster formation rate, ${{\rm d}N(M_{\rm cl}) \over {\rm d}t}
= S \times M_{\rm cl}^{-\alpha}$, is constant;

\item the slope $\alpha$ of the cluster IMF is constant with $N(M_{\rm
cl}) {\rm d}M \propto M_{\rm cl}^{-\alpha} {\rm d}M$;

\item stellar evolution causes clusters to fade as $F_{\lambda} \sim
t^{-\zeta}$, as predicted by cluster evolution models; and

\item the cluster disruption time depends on their initial mass as
$t_{\rm dis} = \tdisref \times (M_{\rm cl} / 10^4 M_\odot)^\gamma$,
where $\tdisref$ is the disruption time-scale of a cluster with $M_{\rm
cl} = 10^4 M_\odot$.  It is well-established, however, that the
disruption time-scale does not only depend on mass, but also on the
initial cluster density and internal velocity dispersion (e.g., Spitzer
1957, Chernoff \& Weinberg 1990, de la Fuente Marcos 1997, Portegies
Zwart et al.  2001).  Following the approach adopted in Paper I,
however, we point out that if clusters are approximately in pressure
equilibrium with their environment, we can expect the density of all
clusters in a limited volume of a galaxy to be roughly similar, so that
their disruption time-scale will predominantly depend on their (initial)
mass (with the exception of clusters on highly-eccentric orbits).  In
the opposite case that the initial cluster density $\rho$ depends on
their mass $M$ in a power-law fashion, e.g., $\rho \sim M^x$ with $x$
being the (arbitrary) power-law exponent, the disruption time-scale will
also depend on mass if $t_{\rm dis} \sim M^a \rho^b$ (Paper I). 

\end{enumerate}
then it can be shown easily that the age distribution of the observed
cluster population will obey the following approximate power-law
behaviours (see Paper I):

\begin{itemize}

\item ${\rm d} N_{\rm cl} / {\rm d} t \propto t^{\zeta(1-\alpha)}$ 
for young clusters due to fading;

\item ${\rm d} N_{\rm cl} / {\rm d} t \propto t^{(1-\alpha)/\gamma}$ 
for old clusters due to disruption.

\end{itemize}
Similarly the mass spectrum of the observed clusters will be 

\begin{itemize}

\item ${\rm d} N_{\rm cl} / {\rm d} M_{\rm cl} \propto M_{\rm
cl}^{(1/\zeta)-\alpha}$ for low-mass clusters due to fading;

\item ${\rm d} N_{\rm cl} / {\rm d} M_{\rm cl} \propto M_{\rm 
cl}^{\gamma-\alpha}$ for high-mass clusters due to disruption.

\end{itemize}

So both distributions will show a double power law with slopes
determined by $\alpha$, $\zeta$ and $\gamma$.  The crossing points are
determined by the cluster formation rate and the characteristic
disruption time-scale of a cluster with an initial mass of $10^4
M_\odot$, $\tdisref$. 

In Paper I we showed that the observed age and mass distributions of the
star cluster systems in four well-studied galaxies indeed show the
predicted double power-law behaviour.  From the analysis of these
observed distributions Boutloukos \& Lamers (2002) showed that the value
of $\gamma$ is approximately constant, $(\gamma = 0.62 \pm 0.06)$, under
the very different environmental conditions in the four galaxies, but
the characteristic disruption time-scales differ significantly from
galaxy to galaxy. 

For the clusters in M82 B the situation is more complex, because
the cluster formation rate was certainly not constant. In fact the 
distribution in Fig. \ref{rates.fig}c suggests that we can distinguish
three phases, which we will discuss separately.

We will assume that the cluster formation rate has been constant within
each phase, but may differ strongly among the phases.  We adopt a
cluster IMF of slope $\alpha=2.0$ (Harris \& Pudritz 1994; McLaughlin \&
Pudritz 1996; Elmegreen \& Efremov 1997; Zhang \& Fall 1999; Bik et al. 
2002).  We will also adopt a slope $\zeta=0.648$ for the evolutionary
fading of clusters in the {\it V} band, derived from the Starburst99
cluster models (see also Paper I).  For the mass scaling of the
disruption time-scale we adopt $\gamma=0.62$ (Paper I). 

\begin{enumerate}

\item {\it The pre-burst phase, at $\log(t/{\rm yr})\ge 9.4$.} The age
distribution of this earliest phase is represented by only three
age-bins of high age at $\log(t / {\rm yr})\ge 9.4$.  The predicted
slope is $(1-\alpha)/\gamma=-1.61$.  We have fitted a straight line
through the last three age-bins with this slope.  We see that the
predicted slope matches the observations within the uncertainties. 

\item {\it The burst phase, in the approximate time interval $8.4 < \log
(t/{\rm yr}) \le 9.4$.} Its distribution can be fitted with a line of
the same predicted slope of $-1.61$.  This is also shown in Fig. 
\ref{rates.fig}c.  The predicted slope fits the observations very well. 
We see that this disruption line of the burst-phase is higher than in
the pre-burst phase by ($0.3 \pm 0.1$) dex.  This implies that the
cluster formation rate during the burst was approximately a factor of 2
higher than before.  However, this factor depends on the assumed
duration of the burst.  The duration of the burst may have been shorter
than the value of $\Delta \log (t)=1.0$ suggested by the figure, because
uncertainties in the derived cluster ages may have broadened the burst
peak.  If the duration of the burst was a factor $X$ shorter than we
adopted, then the overall cluster formation rate during the burst was a
factor $X$ higher than we estimated. 

\item {\it The post-burst phase, at $\log (t/{\rm yr}) < 8.4$.} The age
distribution of this most recent, post-burst phase extends from $6.0 \le
\log (t/{\rm yr}) \le 8.4$.  Unless the disruption time was extremely
short with $\tdisref$ of order a few Myr (which is contradicted by the
analysis of the mass distribution in the burst, see Sect. 
\ref{sec:burstmass}), the age distribution of the youngest bins is
governed by the fading of the clusters and the detection limit.  Despite
the rather large uncertainties for young ages, $\log (t / {\rm yr}) <
7.0$, we find a best-fitting slope $\zeta(1-\alpha) = -0.57 \pm 0.05$,
so that $\zeta = 0.57 \pm 0.05$ (formal uncertainty, not including the
large error bars) for a mass-function slope $\alpha = 2$ (dotted line in
Fig.  \ref{rates.fig}c).  This is, within the rather large
uncertainties, similar to the value of $\zeta \simeq 0.65$ expected from
the Starburst99 and BC00 models.  Therefore we have assumed a fading
line with the predicted slope $\zeta(1-\alpha)=-0.65$ for the youngest
three age bins.  The age bins in the range of $7.0 \le \log(t / {\rm
yr}) \le 8.4$ fall below the extrapolated fading line.  The difference
is about one dex at $\log(t / {\rm yr}) \simeq 8$, although the number
of observed clusters is small.  This suggest that the number of clusters
for ages $t \gtrsim 10^7$ yr is already affected by disruption.  We can
fit a disruption line with the predicted slope of $-1.61$ through the
data points, despite their rather large uncertainties.  We see that the
best-fitting line is displaced by ($-1.6 \pm 0.1 $) dex compared to the
pre-burst disruption line.  This shows that the cluster formation rate
in the post-burst phase was approximately a factor of 40 smaller than
during the pre-burst phase. 

\end{enumerate}
 
We conclude from this analysis that the cluster formation rate during
the burst was at least a factor of 2 higher than during the pre-burst
phase (depending on the duration of the burst), and that the cluster
formation rate was a factor of $\sim 40$ smaller after the burst than
before the burst.  This is not surprising because the intense episode of
cluster formation during the burst will have consumed a large fraction
of the available number of molecular clouds, leaving little material for
cluster formation at later times.  In fact, it is likely that starbursts
are strongly self-limited, or quenched, by supernova-driven outflows,
which remove the remaining cool gas from the immediate starburst region
(e.g., Chevalier \& Clegg 1985, Doane \& Mathews 1993).  The remarkable
minor-axis wind in M82 is a dramatic example of this process.  However,
the disturbed conditions near an early burst may discourage re-ignition
at the same site when cool gas inflows resume, shifting the location of
active star formation (e.g., dGOG). 

\section{The mass distribution of clusters formed in the burst.}
\label{sec:burstmass}

\begin{figure}
\psfig{figure=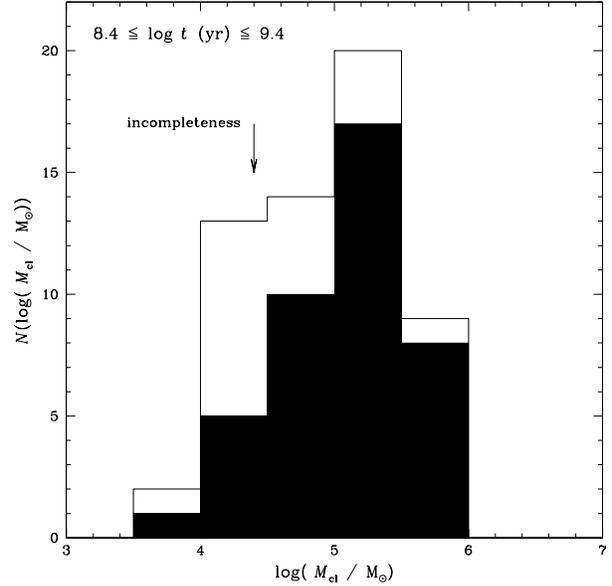,width=8.5cm}
\caption{\label{burstmass.fig}Mass distribution of the clusters formed
in the burst of cluster formation, $8.4 \le \log (t / {\rm yr}) \le
9.4$.  The shaded histograms correspond to the clusters with ($\log(
{\rm Age[max]} ) - \log( {\rm Age[min]} ) \le 1.0$; the open histograms
represent the entire cluster sample in this age range.  The arrow
indicates the mass where the onset of significant selection effects, and
therefore of sample incompleteness, is expected to occur, based on the
selection limit in Fig.  \ref{agemass.fig}.  The expected effect of the
$A_V \lesssim 0.2$ mag extinction for the clusters with well-determined
ages is a shift in mass towards higher masses of $\Delta \log(M_{\rm
cl}/M_\odot) \lesssim 0.08$, which implies that the observed turnover is
not a spurious effect due to varying extinction.}
\end{figure}

In this section we analyse the mass distribution of the clusters that
were formed during the burst of cluster formation, roughly defined to
have occurred in the age range $8.4 \le \log (t/{\rm yr}) \le 9.4$ (see
the vertical dashed lines in Fig.  \ref{rates.fig}c).  In Fig. 
\ref{burstmass.fig} we show their mass distribution.  Although the M82 B
cluster system does not contain large numbers of very massive clusters,
its mass distributions shows a clear increase in cluster numbers from
the high-mass end towards lower masses, for $\log M_{\rm cl}/M_\odot
\gtrsim 5$.  For lower masses, however, the number of clusters decreases
rapidly.  If the mass spectrum of young, newly formed clusters resembles
a power-law distribution, this turn-over at $\log M_{\rm cl}/M_\odot
\simeq 5$ is not expected, unless disruption effects have preferentially
removed the lower-mass clusters from our magnitude-limited sample. 

In dGOG we concluded that the distribution of cluster luminosities (and
therefore the equivalent mass distribution) in M82 B -- corrected to a
common age of 50 Myr -- shows a broader and flatter mass distribution
than typical for young cluster systems, although this distribution is
subject to strong selection effects.  For the small age range considered
here, the selection limit in observable mass at $\log (M/\Msun) \simeq
4.2$ imposed by the brightness limit at $V = 22.5$ mag (Fig. 
\ref{agemass.fig}) occurs at almost an order of magnitude lower masses
than the turn-over mass in Fig.  \ref{burstmass.fig}, at $\log( M_{\rm
cl}/M_\odot) \simeq 5$.  The arrow in Fig.  \ref{burstmass.fig}
indicates the mass where the onset of significant selection effects, and
therefore of sample incompleteness, is expected to occur, based on the
selection limit in the (age vs.  mass) plane shown in Fig. 
\ref{agemass.fig}.  The expected effect of the $A_V \lesssim 0.2$ mag
extinction for the clusters with well-determined ages is a marginal
shift in mass towards higher masses of $\Delta \log(M_{\rm cl}/M_\odot)
\lesssim 0.08$.  This implies that the observed turnover is not a
spurious effect due to varying extinction.  In addition, we do not
observe a systematic trend between cluster mass and extinction, derived
from the SED fitting, which would be expected if the higher-mass
clusters (and therefore brighter at the same age) were located slightly
deeper into M82 B, while the fainter lower-mass clusters were only
observed near the very surface of the region. 

Therefore, we conclude that if the initial mass spectrum of the M82 B
clusters formed in the burst of cluster formation resembled a power-law
distribution down to the low-mass selection limits, cluster disruption
must have transformed this distribution on a time-scale of $\lesssim
10^9$ yrs into a distribution resembling a log-normal or Gaussian
distribution. 

We now compare the observed mass distribution of the clusters formed in
the burst with model predictions.  These predictions are based on the
assumption that the clusters were formed with a cluster IMF of slope
$\alpha=2$, and that the distribution has been modified subsequently by
cluster disruption.  For the cluster disruption we adopt the same
relation $t_{\rm dis}= \tdisref (M/10^4 \Msun)^{\gamma}$ with
$\gamma=0.62$ (Eq.  (\ref{eq:tdis})), as found in Paper I for four
galaxies, and used in the previous section.  We can then model the mass
distribution of the observed burst clusters with only one free
parameter, i.e.  the disruption time $\tdisref$ of a cluster with an
initial mass of $10^4 \Msun$. 

The decreasing mass of a cluster with time is given by\footnote{Recent
simulations, based on both Monte Carlo realisations (e.g., Giersz 2001,
his Fig.  2) and {\it N}-body modeling (e.g., Portegies Zwart et al. 
2002, their Fig.  2; S.J.  Aarseth, priv.  comm.), show that this is a
close approximation to real cluster evolution, provided that the initial
cluster has reached an equilibrium state.}

\begin{equation}
{{\rm d}M \over {\rm d}t} \simeq -{M \over \tdis} =-{M \over {\tdisref (M/10^4
\Msun)^{\gamma}}} \qquad ,
\label{eq:dmdt}
\end{equation}
so that

\begin{equation}
M(M_i,t)= ( M_i^{\gamma}- B t)^{1/\gamma} \qquad ,
\label{eq:mimt}
\end{equation}
where $M$ and $M_i$ are, respectively, the present and the initial mass
of the cluster, both in units of $\Msun$, and $B=\gamma
10^{4\gamma}/\tdisref$. 

Suppose that clusters form in a burst as

\begin{equation}
{\rm d} N(M_i,t)=S(t) M_i^{-\alpha} {\rm d}M_i {\rm d}t
\label{eq:CIMF}
\end{equation}
over the mass range $M_{\rm min} < M_i < M_{\rm max}$, and governed by a
formation rate $S(t)$ in units of $\Msun^{\alpha-1}$ yr$^{-1}$.  We
assume an instantaneous burst of cluster formation, so that $S(t)$ is a
$\delta$ function at age $\tburst = 10^9$ yr.  We adopt $M_{\rm max}=3
\times 10^6 \Msun$ (i.e., $\log M_{\rm max}/M_\odot = 6.5$, one mass bin
beyond the maximum mass of clusters formed in the burst, thus allowing
for some disruption effects) and $M_{\rm min}=10^2 \Msun$, but this
lower limit is not important since it is below the detection limit of
$\sim 10^4$ \Msun\ for the burst clusters (see Fig.  2).  The function
$S(t)$ is related to the total number of clusters formed per unit time
as

\begin{equation}
S(t)={{\rm d}N_{\rm tot} \over {\rm d}t} (\alpha-1) 
\Bigl( M_{\rm min}^{1-\alpha}-M_{\rm max}^{1-\alpha}\Bigr)^{-1} 
\label{eq:St}
\end{equation}
where d$N_{\rm tot}/{\rm d}t$ is the total number of clusters formed per
year.  It is easy to show that the present mass distribution of clusters
formed in an instantaneous burst of age \tburst\ can be written as

\begin{eqnarray}
N(M){\rm d}M &=& S(t) \biggl(1+\gamma\left(\frac{M}{10^4
      \Msun}\right)^{\gamma} \frac{\tburst}{\tdisref}
      \biggr)^{(1-\alpha-\gamma)/\gamma} \nonumber \\
      & & \left(\frac{M}{\Msun}\right)^{-\alpha} {\rm d}M
\label{eq:NM}
\end{eqnarray}
for masses above the detection limit and smaller than $(M_{\rm
max}^{\gamma}-B \tburst)^{1/\gamma}$.  We see that the initial mass
distribution, d$N(M_i,t)= S(t) M_i^{-\alpha}$, has been modified by a
factor that depends on the ratio $\tburst/ \tdisref$, for given values
of $\alpha=2.0$ and $\gamma=0.62$. 

\begin{figure}
\psfig{figure=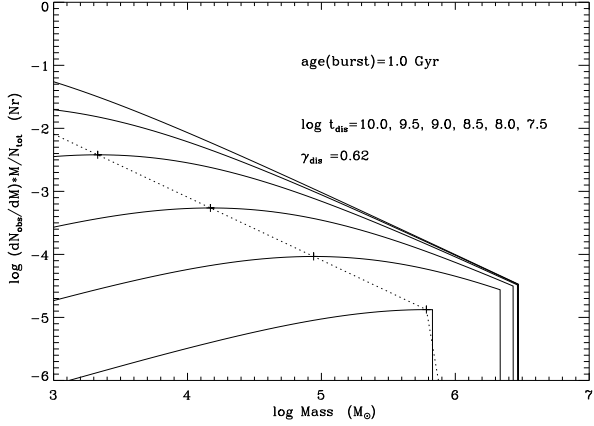,width=8.5cm}
\caption{\label{predmass.fig}Predicted mass distributions in terms of
$\log ({\rm d} N/{\rm d} M) \times (M/N_{\rm tot}$ (equivalent to the
logarithm of the number of clusters per mass bin) for an instantaneous
burst of cluster formation that occurred 1 Gyr ago.  From top to bottom,
the curves represent the (normalised) mass distributions for
characteristic cluster disruption times $\log(\tdisref / {\rm yr}) =
10.0, 9.5, 9.0, 8.5, 8.0$, and 7.5.  Notice the dependence of both the
turnover mass, indicated by the dotted line, and the maximum mass on
$\tdisref$.}
\end{figure}

We have calculated the predicted mass distribution, using Eq. 
(\ref{eq:NM}), for a burst age of 1 Gyr and for several values of the
cluster disruption time-scale $\tdisref$, see Fig.  \ref{predmass.fig}. 
To match the observed peak in the cluster mass distribution, a
characteristic disruption time-scale $\log (\tdisref/{\rm yr})$ between
7.5 and 8.0 is required, with the most likely value closer to $\log
(\tdisref{\rm yr}) = 8.0$. 

\begin{figure}
\psfig{figure=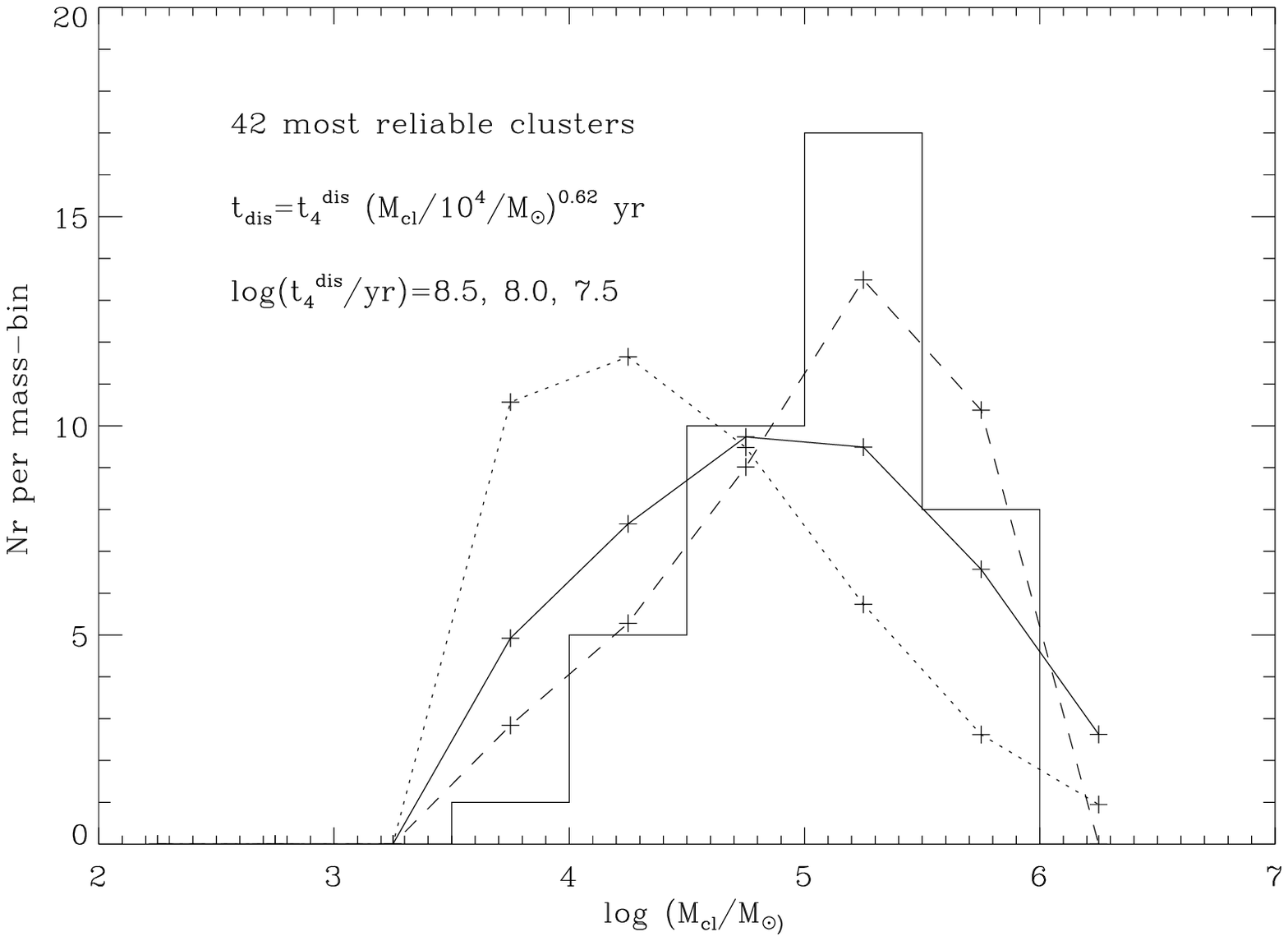,width=8.5cm}
\psfig{figure=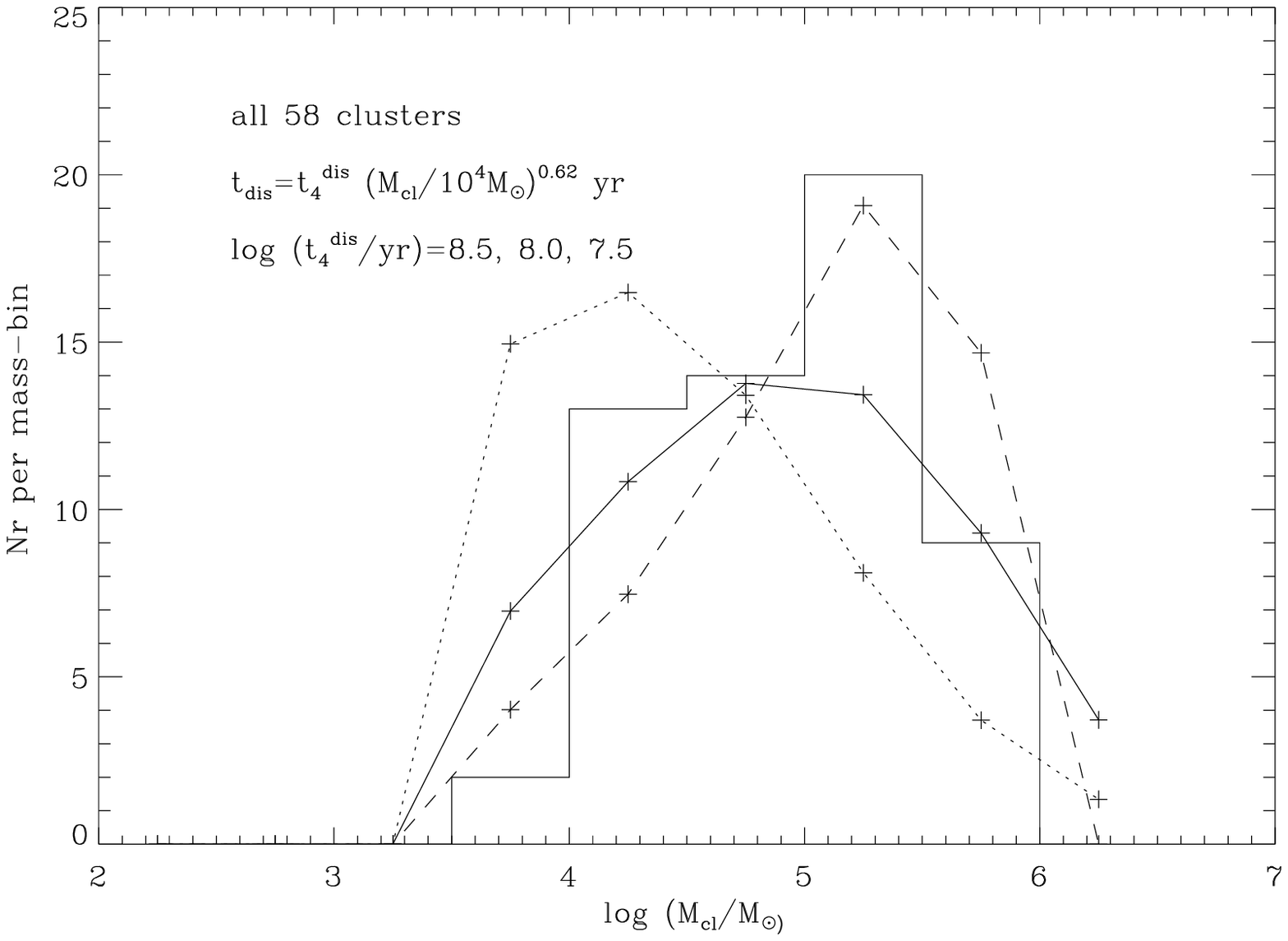,width=8.5cm}
\caption{\label{burstfit.fig} Comparison between the observed mass
distribution of the clusters formed in the burst of cluster formation
and the predicted distribution, based on a cluster IMF with slope
$\alpha=2$ and cluster disruption with time-scales of $t_{\rm dis}=
\tdisref (M/10^4 \Msun)^{0.62}$.  {\it (a)} Cluster sample with the most
accurate masses and ages.  {\it (b)} Full sample.  The predicted number
of clusters in each mass bin, normalised to the total number of observed
clusters, are shown by the dotted, full and dashed lines for
characteristic disruption time-scales $\log (\tdisref/{\rm yr}) = 8.5, \,
8.0$, and 7.5, respectively.}
\end{figure}

We then calculated the expected number of clusters in the same age bins
as observed (Fig.  \ref{burstmass.fig}) and normalised the distribution
to the observed total number of clusters.  The results are shown in Fig. 
\ref{burstfit.fig}a for the subsample of the 42 clusters with the most
accurately determined ages, and in panel b for the full sample of 58
clusters.  The location of the predicted peak is very sensitive to the
value of $\tdisref/\tburst$, which follows from Eq.  (\ref{eq:NM}), and
is shown in Fig.  \ref{predmass.fig}.  We found the best fit for a
cluster disruption time of $\tdisref \simeq 3 \times 10^{-2} \tburst$,
with an uncertainty of approximately a factor of two.  This corresponds
to $\tburst=3 \times 10^7$ yr if $\tburst=1$ Gyr.  We conclude that the
mass distribution of the clusters formed during the burst can be
explained by an initial cluster IMF with a slope of $\alpha=2$, an upper
limit to their mass of $\sim 3\times 10^6$ \Msun\ and cluster disruption
with a time-scale given by Eq.  (\ref{eq:tdis}), $\tdisref \simeq 3
\times 10^7$ yr if the age of the burst is $\tburst = 1 \times 10^9$ yr. 
The uncertainties in both the observed and predicted {\it numbers} of
clusters of a given mass are governed by identical Poisson-type
statistical errors.  The uncertainties in the mass and age scales, on
the other hand, are dominated by systematic uncertainties introduced by
the adopted stellar evolutionary synthesis models, and -- to some extent
-- by the specific extinction law adopted, although the {\it relative}
mass and age distributions are robust.

\section{Discussion} 
\label{discussion.sec}

\subsection{The cluster disruption time-scale}
\label{timescales.sec}

We have derived the characteristic cluster disruption time-scale from
the mass distribution of the clusters formed during the burst.  In
principle this disruption time-scale could also be derived from the the
age distributions of the clusters formed during the post-burst phase,
shown in Fig.  \ref{rates.fig}c.  However the number of observed young
clusters is very small, only a few per age bin so that the crossing
point between the (dotted) fading line and the (solid) disruption line
is very uncertain.  However, the crossing point suggested in Fig. 
\ref{rates.fig}c agrees approximately with that expected for the
disruption time-scale of $\tdisref \simeq 3 \times 10^7$ yr derived from
the clusters formed during the burst (see the description of the method
in Paper I). 

In Paper I we have shown that the disruption time-scale can also be
derived from the mass distribution of clusters in a magnitude limited
sample if the cluster formation rate is constant.  This last assumption
is certainly not justified for M82 B.  For a strongly variable cluster
formation rate, as in M82 B, the mass distribution is heavily affected
such that the crossing point between the power-law fit to the
distribution of low mass clusters (the fading line) and the power-law
fit to the distribution of the massive clusters (the disruption line)
does no longer represent the value of $\tdisref$.  Therefore, we could
only use the mass distribution of the clusters formed during the burst
to derived the disruption time-scale. 

An initial comparison among the characteristic cluster disruption
time-scales found in Paper I reveals large differences.  However, the
characteristic cluster disruption time-scale for the clusters in region
B of its low-mass host galaxy M82 is, within the uncertainties,
comparable to that in the dense centre of the massive grand-design
spiral M51.  Lamers \& Portegies Zwart (in prep.) are currently
analysing the cluster disruption time-scales derived for the galaxies
studied in Paper I and in this paper, using {\it N}-body simulations. 
Here, we will therefore simply explore whether this similarity between
$t^{\rm dis}_4$ in the centre of M51 and in M82 B can be understood from
a comparison of the ambient density in both regions. 

In order to estimate the density in M82 B, we determined its total {\it
V}-band luminosity, after correcting for a nominal extinction throughout
the region of $A_V \sim 0.5$ mag, $V_{\rm tot} \sim 9.5$ from our {\sl
HST} observations (dGOG), so that $M_{V, {\rm tot}} = -18.3$, $L_{V,
{\rm tot}} \simeq 1.8 \times 10^9 L_{V,\odot}$.  Spectral synthesis of
region B suggests that it has a {\it V}-band mass-to-light ratio, $M/L_V
\sim 0.5 - 1.0$ (R.W.  O'Connell, priv.  comm.), as expected from the
Starburst99 models for a stellar population of several 100 Myr, so that
the total mass contained in the objects providing most of the {\it
V}-band luminosity is $M \sim 1.35 \times 10^9 M_\odot$.  If we now
assume that the volume occupied by these objects is roughly similar to a
sphere with a radius determined by the radius of the area in which we
measured the total flux, we derive an average density for M82 B of
$\langle \rho \rangle \sim 2.5 M_\odot {\rm pc}^{-3}$, or $\log \langle
\rho \rangle (M_\odot {\rm pc}^{-3}) \sim 0.4$.  Note, however, that we
have introduced large uncertainties by adopting the above assumptions,
in particular because we have assumed to have sampled the entire volume
of the ``spherical region'' M82 B.  This implies that the region has
been assumed transparent, as opposed to the conclusion in dGOG that the
M82 B cluster sample contains likely only the subset of M82 B clusters
on the surface of the ``sphere''.  The implications of our assumptions
are therefore that we expect to have missed a large number of clusters
present in the interior of M82 B.  For these missed clusters in the
interior we expect the disruption time to be shorter because of the
higher densities predicted there.

For the interstellar medium in the centre of M51, Lamers \& Portegies
Zwart (in prep.) derive a mean density of $\langle \rho \rangle \sim
0.60 M_\odot {\rm pc}^{-3}$, based on column density estimates by
Athanassoula et al.  (1987) and similar geometrical arguments as used
above for M82 B.  Within the large uncertainties involved in such
back-of-the-envelope approximations, these two estimates of the mean
density in the centre of M51 and in M82 B are remarkably similar, within
an order of magnitude, as are their characteristic cluster disruption
time-scales.  Thus, we conclude that, although M82 as such is a small,
low-mass irregular galaxy, its fossil starburst region B has achieved a
similarly dense interstellar medium as the centre of M51, so that
similar cluster disruption time-scales are not {\it a priori} ruled out. 

\subsection{The cluster formation rate}

We have derived the cluster formation history of M82 B from the age
distribution shown in Fig.  \ref{rates.fig}c.  The age distribution
itself, Fig.  \ref{rates.fig}a already shows a very strong peak around
$10^9$ years.  Part of the steep increase between $8 < \log(t/{\rm yr})
< 9$ is owing to the use of logarithmic age bins.  Therefore we have
transformed the age distribution into the formation history of the
observed clusters, Fig \ref{rates.fig}c.  The general decrease with age
clearly shows the effect of cluster disruption.  Disruption of clusters
formed at a constant formation rate and with a mass-dependent disruption
time-scale will result in a power-law decrease of the observed cluster
formation rate (Paper I), roughly in agreement with the observations. 
The deviations from the power law, e.g., the peak in the observed
formation history around 1 Gyr, reflect changes in the real formation
history, or in the disruption time. 

For magnitude-limited cluster samples with a constant cluster formation
rate and a constant mass-dependent disruption time-scale, the disruption
time and its dependence on mass can be derived from the age
distribution.  Because of the non-constant cluster formation rate and
the small number of clusters, such an analysis is not possible here. 
Therefore we opted for the reasonable alternative of adopting the mass
dependence of the disruption time-scale, i.e., the value of $\gamma
\simeq 0.62$ in Eq.  (\ref{eq:tdis}), which seems to be a universal
value found in four galaxies with very different conditions (Paper I). 
We also assumed that the scaling factor $\tdisref$ of the disruption
time-scale is constant.  With these two assumptions, we derived the
ratios between the real cluster formation rates in the pre-burst phase
($\log(t/{\rm yr})\ge 9.4$), the burst-phase ($8.4 < \log(t/{\rm yr}) <
9.4$) and the post-burst phase ($\log(t/{\rm yr}) \le 8.4$ of, roughly,
$1:2:{1 \over 40}$.  The formation rate during the burst may have been
higher if the actual duration of the burst was shorter than adopted. 

Alternatively, we could have adopted a constant cluster formation rate,
but wildly variable cluster disruption time-scales.  In that case the
cluster disruption time-scale should have been much longer (i.e.  much
slower disruption) during the burst than before the burst, and again
much shorter (faster disruption) after the burst.  We think that this
alternative is very unlikely because (a) disruption is a slow process
that occurs over an extended period of time, so large changes are not to
be expected, (b) there is no obvious physical process that would produce
such large changes and (c) the alternative of a variable cluster
formation rate is much more likely, because this effect is observed in
several interacting galaxies and in M82 regions A, C and E (dGOG). 
Therefore, we are confident that the changes in the apparent formation
rates of the observed clusters are due to changes in the real cluster
formation rate. 

\section{Summary and Conclusions}
\label{summary.sec}

In this paper, we have reanalysed the previously published optical and
near-infrared {\sl HST} photometry of the star clusters in M82's fossil
starburst region B (dGOG) to obtain improved individual age and mass
estimates.  We have also extended this study to obtain estimates for the
cluster formation history and include the importance of cluster
disruption.  Our main results and conclusions can be summarised as
follows:

\begin{enumerate}

\item Our new age estimates, based on improved fitting methods, confirm
the peak in the age histogram attributed to the last tidal encounter
with M81 by dGOG.  For the subsample of clusters with well-determined
ages we find a peak formation epoch at $\log( t_{\rm peak} / {\rm yr} )
= 9.04 \; (t_{\rm peak} \simeq 1.10$ Gyr), with a Gaussian $\sigma$ of
$\Delta \log( t_{\rm width}) = 0.27$, corresponding to a FWHM of $\Delta
\log( t_{\rm width}) = 0.64$.  The corresponding numbers for the full
sample are $\log( t_{\rm peak} / {\rm yr} ) = 8.97 \; (t_{\rm peak}
\simeq 0.93$ Gyr) for the peak of cluster formation, and for $\sigma$,
$\Delta\log( t_{\rm width}) = 0.35$ (FWHM, $\Delta \log( t_{\rm width})
= 0.82$).  In dGOG (see also de Grijs 2001), we concluded that there is
a strong peak of cluster formation at $\sim 650$ Myr ago, which have
formed over a period of $\sim 500$ Myr, but very few clusters are
younger than 300 Myr ($\log( t / {\rm yr} ) \simeq 8.5$).  Our new age
estimates date the event triggering the starburst to be slightly older. 
Our estimate of the duration of the peak is an upper limit because the
observed width of the peak in the age histogram may have been broadened
by uncertainties in the derived cluster ages. 

\item The improved mass estimates confirm that the (initial) masses of
the young clusters in M82 B with $V \le 22.5$ mag are mostly in the
range $10^4 - 10^6 M_\odot$, with a median of $10^5 M_\odot$ (dGOG).  We
find that the mean mass of our M82 B cluster sample is $\log( M_{\rm cl}
/ M_\odot ) = 5.03$ and 4.88 for the subsample with well-determined ages
and the full sample, respectively, corresponding to $M_{\rm cl} = 1.1
\times 10^5$ and $0.8 \times 10^5 M_\odot$, respectively.  If the
initial mass spectrum of the clusters formed at the burst epoch was a
power law, disruption effects have transformed it into a broader and
flatter distribution on time-scales of $\lesssim 1$ Gyr. 

\item The (apparent) formation history of the observed clusters shows a
gradual increase to younger ages, reaching a maximum at the present
time.  This shows that cluster disruption must have removed a large
fraction of the older clusters.  Adopting the expression for the cluster
disruption time, $\tdis(M)= \tdisref (M/10^4 \Msun)^{\gamma}$ with
$\gamma \simeq 0.62$, that was derived for four galaxies characterized
by very different conditions (Paper I), we found that ratios between the
real cluster formation rates in the pre-burst phase ($\log(t/{\rm
yr})\ge 9.4$), the burst-phase ($8.4 < \log(t/{\rm yr}) < 9.4$) and the
post-burst phase ($\log(t/{\rm yr}) \le 8.4$) are roughly $1:2:{1 \over
40}$.  The formation rate during the burst may have been higher if the
actual duration of the burst was shorter than adopted.  We see that the
cluster formation rate in the post-burst phase is much smaller than in
the pre-burst phase, because the burst has consumed a large fraction of
the available molecular clouds, leaving little material for the cluster
formation in the post-burst phase. 

\item The mass distribution of the clusters formed during the burst
shows a turnover at $\log(M_{\rm cl}/M_\odot) \simeq 5.2$.  This
turnover is not due to selection effects, because the magnitude limit of
our sample would produce a turnover near $\log(M_{\rm cl}/M_\odot)
\simeq 4.5$.  The observed distribution can be explained by cluster
formation with an initial power-law mass function of exponent $\alpha=2$
up to a maximum cluster mass of $M_{\rm max}=3 \times 10^6 \Msun$, and
cluster disruption given by Eq.  (\ref{eq:tdis}) with the adopted value
of $\gamma=0.62$ with only one free fitting parameter: $\tdisref /
\tburst \simeq 3 \times 10^{-2}$.  For a burst age of $1 \times
10^9$ yr, we find that $\tdisref \simeq 3 \times 10^7$ years, with an
uncertainty of a factor of two. 

\item The time-scale of $\tdisref \sim 30$ Myr is much shorter than
derived in Paper I for any of the SMC ($\sim 8$ Gyr), the solar
neighbourhood ($\sim 1$ Gyr), M33 ($\sim 0.13$ Gyr) and even shorter
than in the inner spiral arms of M51 ($\sim 40$ Myr).  The
characteristic disruption time-scale in M82 B is the shortest known in
any disc (region of a) galaxy. 

\end{enumerate}

\section*{Acknowledgments} RdeG wishes to thank the Astronomical
Institute of Utrecht University for their hospitality and partial
support on a visit when this work was started.  We acknowledge
interesting and useful discussions with Bob O'Connell and Sverre Aarseth
and helpful suggestions by the anonymous referee.  This research has
made use of NASA's Astrophysics Data System Abstract Service.

\end{document}